\documentclass[aps,prl,reprint,superscriptaddress,longbibliography,floatfix]{revtex4-2}

\usepackage{graphicx}
\usepackage{amsmath}
\usepackage{amssymb}
\usepackage[nopatch=footnote]{microtype}
\usepackage{hyperref}
\hypersetup{colorlinks=true,linkcolor=blue,citecolor=blue,urlcolor=blue}
\begin{document}

\title{Reciprocal asymmetric transmission in self-shadowed metallized gratings}

\author{Can Yesilyurt}
\email{cyesil@nanorescenter.com}
\affiliation{Nanoelectronics Research Center, Istanbul, Turkey}

\date{\today}

\begin{abstract}
A single oblique metallization step gives conventional gratings
one-sided silver decoration, established by electron microscopy and
elemental mapping. Opposite-face reversal measurements yield
finite-aperture red/green/blue transmission ratios up to $1.74$ for
a 1-$\mu$m grating and $1.5$--$1.8$ for a 30-$\mu$m grating, with
reference-normalized transmittances for the latter. Flat-side
coating and reduced silver dose give ratios near unity. Maxwell
calculations on self-shadowed growth profiles reproduce the
directional sign qualitatively and attribute the contrast to
diffraction-channel redistribution. The structure remains
reciprocal, including its normal-incidence zeroth-order
transmission. These results establish directional metallization as
a route to direction-selective optical transmission without
multilayer alignment or overlay lithography.
\end{abstract}

\maketitle

\emph{Introduction.}---Reciprocity permits asymmetric transmission
between restricted sets of optical channels. For a linear,
time-invariant structure composed of reciprocal media, Lorentz
reciprocity relates power-normalized scattering amplitudes for
exchanged input and output channels~\cite{Jalas2013,Shi2015},
\begin{equation}
S^{F}_{mn} = S^{B}_{\tilde n \tilde m},
\label{eq:recip}
\end{equation}
where $F$ and $B$ denote opposite illumination directions and a
tilde denotes the time-reversed modal partner. At normal incidence
the corresponding zeroth-order channels of a reciprocal,
polarization-preserving grating therefore obey
\begin{equation}
T^{F}_{0} = T^{B}_{0} \qquad \text{(normal incidence)}.
\label{eq:T0}
\end{equation}
Directional contrast can nevertheless arise through unequal
coupling to higher diffraction orders and through oblique
illumination components. Finite modal preparation and collection
select that redistribution without violating Eq.~(\ref{eq:recip}).

Polarization-converting structures~\cite{Fedotov2006,Huang2012}
and asymmetric diffractive gratings~\cite{Lockyear2006,
Serebryannikov2009,Ye2010OE,Cakmakyapan2010,Stolarek2013,Zhu2014,
Tang2016,Xu2014,Yao2016} exploit this distinction by arranging
different couplings among accessible optical
channels~\cite{Ansari2023}. For normal-incidence transmission
through polarization-preserving gratings, higher-order channel
opening provides the multichannel basis for directional
asymmetry~\cite{Foteinopoulou2022}. Many implementations use
deliberately patterned asymmetric geometries; an alternative is to
let directional deposition create the asymmetry on an existing
grating.

Oblique-angle deposition shapes growing films by line-of-sight
self-shadowing~\cite{Barranco2016,Hawkeye2007,Flanders1981}.
Coating replicated binary gratings in this way produces asymmetric
blaze profiles and orientation-dependent color
effects~\cite{Lutolf2014,Lutolf2015}, while oblique metallization
of nanogratings has also been used for wire-grid
polarizers~\cite{Chen2007}. Qing \emph{et al.} recently studied
lateral reciprocal asymmetric transmission in directionally
Ag-coated silica nanogratings~\cite{Qing2026}: they compare
opposite in-plane wavevectors from the same half-space. The
opposite-face, or vertical, sample-reversal configuration examined
here is complementary to that lateral geometry.

Here we show that a conventional grating acquires substantial
reciprocal directional transmission through a single oblique
metallization step. Surface and compositional measurements
establish one-sided Ag redistribution, and sample-reversal
measurements on 1-$\mu$m and 30-$\mu$m gratings establish
finite-aperture directional contrast. Flat-side coating and
reduced-dose controls test its structural origin. Maxwell
calculations on self-shadowed growth profiles explain the
substrate-favored sign and identify diffraction-channel
redistribution as the mechanism. This connects a simple
fabrication step to a measured optical response without requiring
multilayer alignment or overlay lithography.

\begin{figure*}[tb]
\includegraphics[width=\textwidth]{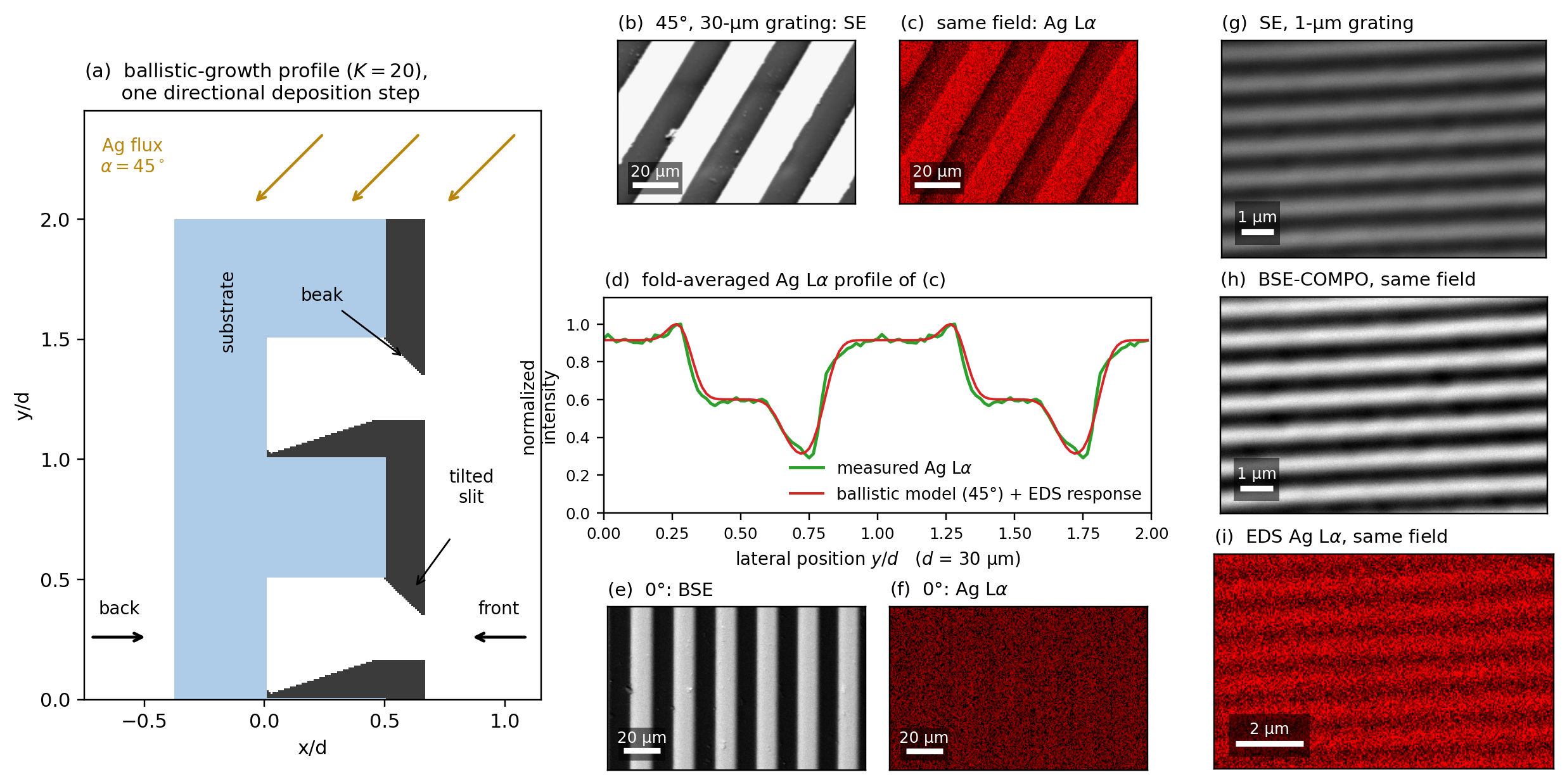}
\caption{\label{fig:fab}
\textbf{Directional metallization: model and structural evidence.}
(a)~Ballistic-growth model ($\alpha = 45^\circ$, $K=20$ growth
steps) of a binary grating (blue) under collimated Ag flux (arrows).
The predicted profile has coated tops and flux-facing sidewalls, bare shadowed
groove floors, and corner overhangs that form tilted slits;
front/back label the two illumination directions.
(b)~Secondary-electron (SE) image of the $45^\circ$-coated
30-$\mu$m grating and (c)~the energy-dispersive x-ray
spectroscopy (EDS) Ag~L$\alpha$ map of the same field: periodic
\emph{one-sided} Ag lanes.
(d)~Fold-averaged Ag~L$\alpha$ lateral profile of (c) versus the
ballistic-growth model passed through a minimal EDS emission
response (generation depth $0.4\,\mu$m). The modeled Ag rim,
shadowed strip, and partially shadowed groove floor account for
the main features of the measured lateral profile.
(e)~Backscattered-electron (BSE) image of a grating metallized at
normal incidence and (f)~its Ag~L$\alpha$ map,
laterally \emph{uniform}: the one-sidedness is created by the
oblique flux.
(g,h,i)~SE, BSE, and EDS Ag~L$\alpha$ images of the \emph{same
field} of the $45^\circ$-coated 1-$\mu$m grating (the sample of
Fig.~\ref{fig:app}).}
\end{figure*}

\emph{Directional metallization and structural evidence.}---The
experimental gratings are metallized at $45^\circ$ with the
deposition azimuth perpendicular to the grating lines. This
macroscopic orientation setting directs the coating onto the
relief without overlay alignment. Electron microscopy and
Ag~L$\alpha$ mapping show periodic one-sided Ag lanes on the
obliquely coated gratings [Fig.~\ref{fig:fab}(b)--(d),(g)--(i)].
The normal-incidence deposition control has a laterally uniform
Ag map [Fig.~\ref{fig:fab}(e,f)], linking the one-sided
redistribution to the oblique flux.

On the 30-$\mu$m grating, secondary-electron (SE) and
backscattered-electron (BSE) contrast maxima occur at phases
differing by $0.34$ of the period. The channels are acquired at
registered scan positions with a calibrated SE/BSE detector pair,
so the displacement separates topographic and compositional
maxima on the same coated line. The fold-averaged Ag~L$\alpha$
lateral profile is reproduced by a ballistic-growth model passed
through an energy-dispersive x-ray spectroscopy (EDS) response
(correlation $0.99$) [Fig.~\ref{fig:fab}(d)]. The correspondence
supports self-shadowing as the origin of the directional Ag
distribution.

The model assumes collimated flux, unity sticking, line-of-sight
occlusion, and cellular growth~\cite{Barranco2016,
Nieuwenhuizen1966,Dirks1977,Tait1993}. At $45^\circ$ it predicts
coated tops and flux-facing sidewalls, shadowed groove floors,
and corner overhangs that form tilted slits
[Fig.~\ref{fig:fab}(a)]. This louver-like geometry provides a
reflective analogue of laminated absorptive micro-louver
films~\cite{Takatoh2023}. Surface and compositional imaging
establish one-sided Ag redistribution consistent with this growth
mechanism. Growth parameters and the predicted thickness dependence
are given in the Supplemental Material~\cite{SM}.

The geometric interpretation is broken \emph{interface orientation
symmetry} (IOS): reversing the structure exchanges inequivalent
entry and exit interface orientations. The same geometric idea
appears in ballistic Dirac transport~\cite{YesilyurtJAP2026},
while in the optical grating the periodicity allows momentum
transfer in reciprocal-lattice increments,
\begin{equation}
k_{\parallel,\mathrm{out}} = k_{\parallel,\mathrm{in}} + mG,
\qquad G = 2\pi/d.
\label{eq:modG}
\end{equation}
Here $d$ is the period and $m$ the diffraction order. The
asymmetric profile changes coupling among these channels. IOS
describes that geometry; the quantitative optical interpretation
below follows from Maxwell electromagnetics.

\begin{figure*}[t]
\includegraphics[width=\textwidth]{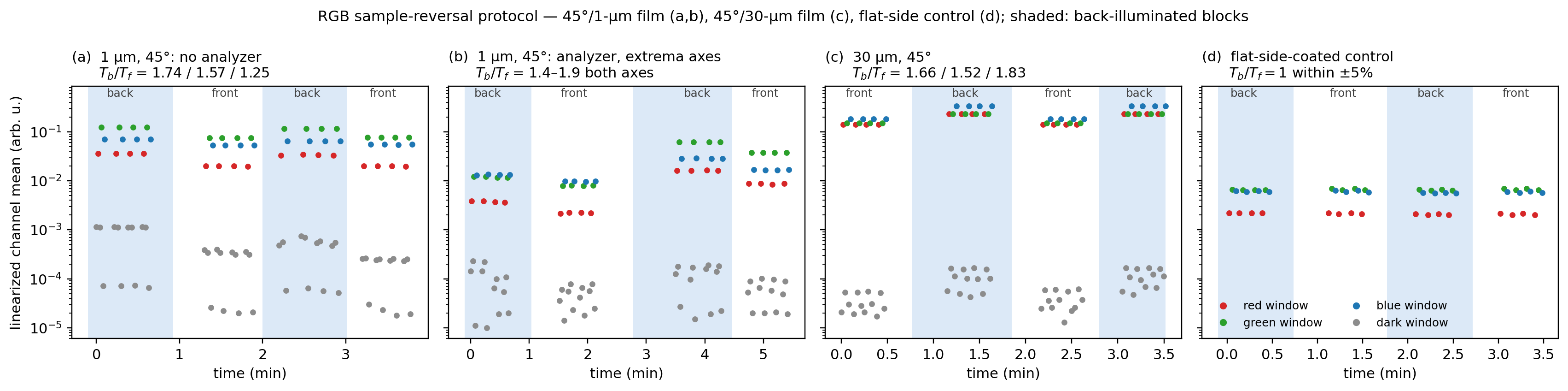}
\caption{\label{fig:app}
\textbf{Measured finite-aperture directional transmission and controls}
(Cr/Ag metallization). Front denotes the coated face toward the
source; back denotes substrate-side illumination. Each panel is
one acquisition session; shaded blocks are back-illuminated,
colored points are individual R/G/B windows, and gray points are
dark windows. Ratios use dark-corrected block means.
(a)~Primary result: the $45^\circ$, 1-$\mu$m grating gives
$C = T_b/T_f = 1.74/1.57/1.25$ (R/G/B).
(b)~Output-analyzer test: contrast persists in both orthogonal
projections ($1.4$--$1.9$).
(c)~Independent large-period result: the $45^\circ$, 30-$\mu$m
grating gives $C = 1.66/1.52/1.83$. A no-sample reference in the
same session gives absolute finite-aperture transmittances
$T_f = 0.14$--$0.19$ and $T_b = 0.24$--$0.34$.
(d)~Flat-side-coated control: the 1-$\mu$m grating stock coated
at $45^\circ$ on its flat rear face, leaving the corrugation bare,
gives $C = 1$ within $\pm 5\%$. Acquisition and normalization
details are given in the Supplemental Material~\cite{SM}.}
\end{figure*}

\emph{Measured finite-aperture directional transmission.}---The
extended red/green/blue (RGB) light-emitting-diode (LED) source
illuminates the sample over a distribution of incident angles,
while a finite collection aperture admits only part of the
transmitted modal distribution. For a specified incident mode,
the collected transmittance is
\begin{equation}
T_D = \frac{1}{P_\mathrm{in}} \int_{\Omega_D}
\frac{dP_\mathrm{out}}{d\Omega}\,d\Omega,
\label{eq:TD}
\end{equation}
where $P_\mathrm{in}$ is incident power and $\Omega_D$ is the
collection solid angle. The experiment measures an incoherent
average of this quantity over incident position, angle, wavelength,
and polarization, weighted by source radiance, detector
responsivity, and the collection transfer function. These fixed
weights define the finite-aperture directional observable.

Within each acquisition session the sample is reversed while
source and detector settings remain fixed. The directional
contrast is $C=T_\mathrm{back}/T_\mathrm{front}$, with front
denoting illumination of the coated face and back denoting
illumination through the substrate. Color windows are bracketed
by dark windows, and each block contains four repeats. Stable
multiplicative gains cancel in the ratio; the spectral and angular
weights remain part of the observable. Within-block scatter is
$\leq 1\%$, which describes technical repeatability, not the full
remounting uncertainty. The geometry, timing, linearization, and
reference normalization are detailed in the Supplemental
Material~\cite{SM}.

For the $45^\circ$-coated 1-$\mu$m grating, the measured contrast
is $C=1.74/1.57/1.25$ (R/G/B), favoring substrate-side
illumination [Fig.~\ref{fig:app}(a)]; the directional difference
is also visible in white light. Substantial asymmetry persists
in both orthogonal output-analyzer projections ($1.4$--$1.9$)
[Fig.~\ref{fig:app}(b)]. This analyzer test establishes persistence
in the measured output projections; it does not measure the
separate incident TE and TM responses.

For the $45^\circ$-coated 30-$\mu$m grating, the ratios are
$C=1.66/1.52/1.83$ [Fig.~\ref{fig:app}(c)]. A no-sample
reference recorded in the same session gives absolute
finite-aperture transmittances $T_f=0.14$--$0.19$ and
$T_b=0.24$--$0.34$. At $d/\lambda\approx46$--$67$ across the
visible, these measurements establish substantial directional
contrast at the sampled colors in the large-period regime.

Two optical controls test the connection to the coating of the
grating relief. Coating the flat rear face of the same 1-$\mu$m
stock at $45^\circ$, while leaving the corrugated face bare,
gives $C=1$ within $\pm5\%$ [Fig.~\ref{fig:app}(d)]. Planar
directional metallization alone is therefore insufficient to
produce the observed contrast in this geometry. A reduced-dose
$45^\circ$ coating on the corrugated face gives
$C=1.02/1.00/0.97$. Together with the normal-incidence morphology
control, these observations support a structural origin tied to
directional Ag redistribution on the relief. The dose dependence
is consistent with the modeled development of overhangs.

The measured asymmetry is consistent with the reciprocal-channel
picture. For $d=1\,\mu$m at normal incidence the $\pm1$ orders
leave at $27^\circ$--$47^\circ$ across the visible. Near-axis LED
elements contribute predominantly the zeroth order, whereas
sufficiently oblique source elements can redirect first orders
into the collection cone. Sample reversal changes coupling among
the admitted channels. Equation~(\ref{eq:recip}) exchanges input
and output channels as well as their propagation directions.
Summing all transmitted orders for one fixed input sums squared
amplitudes down one column of $S^F$; reciprocity relates this to
the corresponding squared-amplitude sum across a row of $S^B$,
not to the column selected by the same nominal incident mode
from the opposite face. These two all-order
transmittances can differ. Complete, equally weighted sums over
reciprocally exchanged input and output channel sets are equal.

\emph{Maxwell interpretation.}---We use two-dimensional
finite-difference time-domain (FDTD) calculations~\cite{Yee1966,Taflove2005}
with the ballistic-growth geometry, periodic lateral boundaries, and
absorbing layers~\cite{Roden2000}. The reference model treats Ag
as a perfect conductor and the dielectric as $n=1.55$; TE
($E\parallel$ lines) and TM ($H\parallel$ lines) are computed
separately at normal incidence. Here $T^{F,B}$ is the transmitted
flux summed over \emph{all} propagating orders for a plane-wave
input, normalized to a vacuum reference. Unpolarized contrasts
are formed from power averages,
$T_\mathrm{unpol}=(T_\mathrm{TE}+T_\mathrm{TM})/2$ in each
direction. Numerical geometry, band averaging, and convergence
checks are given in the Supplemental Material~\cite{SM}.

\begin{figure}[tbp]
\includegraphics[width=8cm]{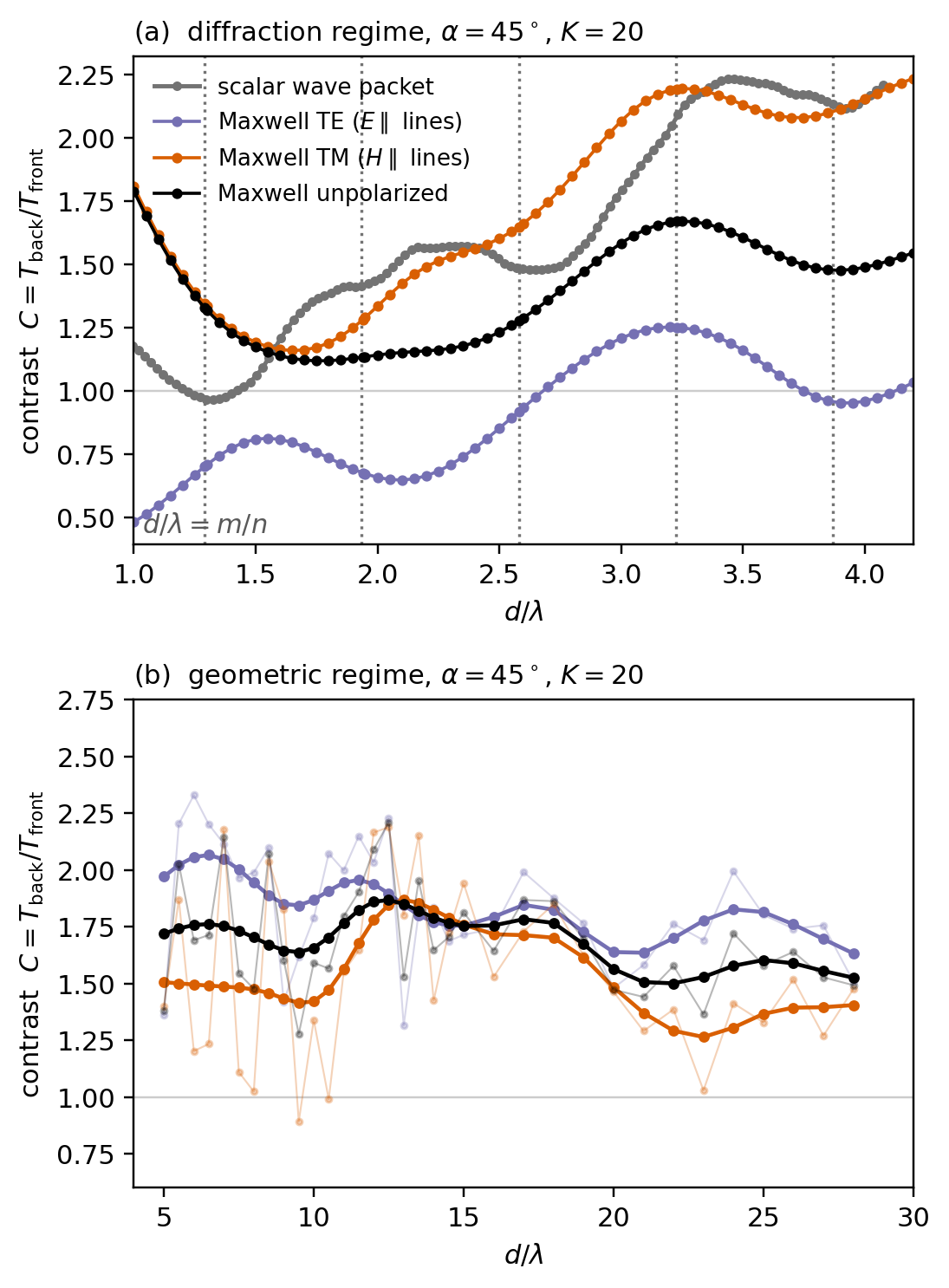}
\caption{\label{fig:maxwell}
\textbf{Calculated directional contrast.} Maxwell FDTD results
for TE ($E\parallel$ lines), TM ($H\parallel$ lines), and
unpolarized illumination (from power-averaged transmittances),
with the scalar wave-packet comparison, for the $45^\circ$,
$K=20$ ballistic-growth profile ($n=1.55$).
(a)~Diffraction regime; dotted vertical lines mark in-medium
Rayleigh thresholds $d/\lambda=m/n$.
(b)~Geometric-regime prediction on $3\times$/$4\times$ upsampled
grids. Bold curves are band averages, light curves show raw grid
points, and TE approaches TM over the modeled range up to
$d/\lambda=28$. Gaussian widths are $\sigma_{d/\lambda}=0.25$
in (a) and $1.2$ in (b). Circles mark computed points; connecting
lines guide the eye.}
\end{figure}

The calculations satisfy Eq.~(\ref{eq:T0}) to numerical
accuracy, while the $\pm1$-order transmittances differ between
the two directions by tens of percent. The modeled bare grating
gives $C=1$ within $4\%$. For the $45^\circ$ louver profile,
the diffraction-regime contrast is mainly carried by TM,
reaching $2.10$ at $d/\lambda=3.6$ (unpolarized $1.56$), while
TE is weak or reversed [Fig.~\ref{fig:maxwell}(a)]. Thus the
model predicts incident-polarization contrasts of approximately
$2$ for $H\parallel$ lines and $1$ for $E\parallel$ lines.
For comparison, the scalar wave-packet
model~\cite{DiracWavepacket,Berry1987} illustrates propagation
through the same asymmetric geometry [Fig.~\ref{fig:maxwell}(a)].
Its boundary conditions differ from electromagnetics, so
quantitative optical predictions use the Maxwell results.

The modeled substrate-favored sign agrees qualitatively with the
experiment. A quantitative comparison would require averaging
computed optical responses over the experimental angular,
spectral, polarization, and collection weights, using the actual
coating geometry.

At larger $d/\lambda$, band-averaged unpolarized calculations
give contrasts of $1.5$--$1.9$ up to $d/\lambda=28$, with TE
approaching TM [Fig.~\ref{fig:maxwell}(b)]. The 30-$\mu$m RGB
ratios have comparable magnitudes at $d/\lambda\approx46$--$67$,
outside the calculated range. In the lossless model, rejected
power is reflected:
$R_\mathrm{front}\approx0.85$ and
$R_\mathrm{back}\approx0.70$ for TM, with $84$--$99\%$ of
reflected power in the specular order.

The optical interpretation is tested against material and
profile idealizations. A Drude-silver calculation retains the
visible-band contrast of the 1-$\mu$m model within $\sim10\%$
of the perfect-conductor values (TM $1.2$--$1.6$ across the
measured channels) and predicts the largest unpolarized contrast
in the red. A shallower profile fitted to the lateral EDS signal
retains the substrate-favored sign at smaller unpolarized
contrast ($1.03$--$1.09$). In the calculated dose series the
unpolarized contrast rises from $\approx0.9$ at
$\tau\approx0.07\,d$ to $1.3$--$2.0$ at
$\tau\approx0.23\,d$, crossing unity near
$\tau\approx0.1\,d$. This sensitivity supports the interpretation
that development of the overhangs controls the contrast, while
showing why a profile inferred from surface data alone does not
fix its magnitude. Toward the subwavelength cutoff the modeled
contrast approaches unity~\cite{Foteinopoulou2022}; sharp
Rayleigh--Wood features in the calculated spectra~\cite{Hessel1965,
Maradudin2016,Sarrazin2003,Treacy2002} remain predictions for
wavelength-resolved testing. Material, profile, and spectral
details are retained in the Supplemental Material~\cite{SM}.

Numerical checks reproduce the analytic slab Fabry--P\'erot
response to $0.6\%$ and conserve energy to
$|T+R-1|\leq0.9\%$ ($1.8\%$ on the geometric-regime grids).
At $d/\lambda=28$, the two finest grids agree within $8\%$
for the band-averaged unpolarized contrast. This discretization
limit applies to the smoothed prediction; the larger differences
at individual spectral points are reported in the Supplemental
Material~\cite{SM}.

Maxwell calculations also predict that deposition angle and dose
tune the magnitude and sign of the directional response; the
complete angular calculation is given in the Supplemental
Material~\cite{SM}.

\emph{Conclusion and outlook.}---A single directional
metallization step produces one-sided Ag redistribution and
substantial opposite-face transmission contrast on conventional
gratings. The 1-$\mu$m and 30-$\mu$m RGB results, output-analyzer
test, and optical controls establish a finite-aperture response
whose sign is consistent with Maxwell calculations of
diffraction-channel redistribution. Broken IOS provides a
geometric interpretation within reciprocal optics.

Further refinement can be organized around direct cross-sectional
structural characterization and angle-, incident-polarization-,
and power-resolved optical measurements. Established grating
replication and oblique-deposition methods~\cite{Lutolf2014,
Lutolf2015} make directional coatings relevant to prospective
glazing, daylighting, and diffuser studies; application-specific
throughput, angular acceptance, and durability would require
evaluation. Directional deposition generates the asymmetric
optical functionality without overlay alignment or an explicitly
aligned asymmetric multilayer. A conventional grating thus
acquires reciprocal direction-selective transmission through a
single self-shadowing metallization step.

\begin{acknowledgments}
This work was funded by the Nanoelectronics Research Center (Nano
Electronics Research M\"uhendislik Ara\c{s}t{\i}rma Geli\c{s}tirme
Ltd.\ \c{S}ti.), Istanbul, Turkey.
\end{acknowledgments}

\emph{Data availability.}---The data that support the findings of
this study are available from the corresponding author upon
reasonable request.

\bibliography{references}

\onecolumngrid
\raggedbottom
\vspace{2\baselineskip}
\setcounter{section}{0}
\setcounter{figure}{0}
\setcounter{equation}{0}
\renewcommand{\thefigure}{S\arabic{figure}}
\renewcommand{\theequation}{S\arabic{equation}}
\renewcommand{\thesection}{S\arabic{section}}

\begin{center}
{\large\bfseries Supplemental Material for ``Reciprocal asymmetric transmission in self-shadowed metallized gratings''}\\[0.6em]
Can Yesilyurt\\
{\itshape Nanoelectronics Research Center, Istanbul, Turkey}
\end{center}
\vspace{0.5em}

\noindent This Supplemental Material describes the measurement geometry and
protocol, the ballistic-growth model and the Maxwell calculation, the
material, profile, and deposition-angle predictions, and the numerical
convergence of the geometric-regime grids. Figures and equations are
numbered S1, S2, \ldots; ``Fig.~1''--``Fig.~3'' refer to the main text.
\suppressfloats[t]

\section{Measurement geometry and protocol}

\begin{figure}[tb]
\includegraphics[width=0.7\textwidth]{fig_s1_schematic.pdf}
\caption{\label{fig:schematic}
\textbf{Directional-transmission measurement.}
(a)~Geometry: the R/G/B LED matrix source illuminates the mounted
sample from below; the transmitted light is collected by a
$3.2\times$ low-aperture objective at $40$-mm working distance,
focused on the sample plane and sampling a $\approx 5$-mm field of
view.
(b)~Sample-reversal protocol: the sample is flipped between
measurement blocks, so that the same incident beam ($k_i$)
illuminates either the coated face (front, top) or the substrate
face (back, bottom); the directional contrast
$C = T_\mathrm{back}/T_\mathrm{front}$ is formed from block means
recorded within a single locked acquisition session.}
\end{figure}

Figure~\ref{fig:schematic} shows the directional-transmission
measurement of the main text. The source is a red/green/blue
(R/G/B) light-emitting-diode (LED) matrix
driven through a fixed color sequence; each color window is
$0.5$~s long, with dark windows interleaved between the colors, and the sequence is
preceded by a fixed color preamble from which the detector
synchronizes optically. The detector is a photosensor operated at
fixed frame rate with focus, exposure, gain, and white balance
converged once and then locked for the entire session; channel
means are formed over a centered region of interest after
sRGB decoding to linear light. In the archived exports, the
channel-specific mean of the three dark windows in each
dark/red/dark/green/dark/blue sequence is subtracted from the
corresponding color-window mean.

The sample is mounted on spacers directly above the LED matrix,
and the transmitted light is collected by a $3.2\times$ objective
($7$-mm clear aperture at $40$-mm working distance, i.e.\ a
geometric acceptance half-angle of at most
$\arctan(3.5/40) \approx 5^\circ$) focused on the sample plane,
sampling a $\approx 5$-mm field of view. Because the LED matrix is spatially extended, emitters
far from the optical axis illuminate the sample obliquely, and
their first diffraction orders can be redirected into the
collection cone ($\sin\theta_\mathrm{out} =
\sin\theta_\mathrm{in} \pm \lambda/d$), so both the zeroth order
(from near-axis emitters) and first orders (from off-axis
emitters) contribute to the collected signal. The measured
quantity is therefore the finite-aperture directional
transmittance defined in the main text, recorded with source
and collection geometry held fixed while the sample is reversed;
the distribution of the collected light among the admitted
diffraction orders is itself direction-dependent and is part of
the measured asymmetry.

One block consists of four consecutive source sequences recorded
with the sample fixed in place; the sample is then flipped
[Fig.~\ref{fig:schematic}(b)] and the next block is recorded,
alternating front and back illumination. All quoted contrasts are
ratios of block means taken within one locked session: identical
source and detector settings across blocks cancel stable
multiplicative gains, and the source spectrum, detector
responsivity, incident-angle distribution, and collection transfer
function enter both block means as the same fixed weights of the
measured ratio. For the principal 1-$\mu$m and 30-$\mu$m datasets,
the within-block sample standard deviation divided by the block
mean is below $1\%$ in every color; this describes technical
repeatability within a fixed mounting.

Reversal/remounting can change alignment and shift the
$\approx5$-mm sampled region across a spatially nonuniform
coating. Within-block scatter therefore does not quantify the
combined effects of remounting and spatial sampling.

For the 30-$\mu$m grating, a no-sample reference recorded in the
same session normalizes the dark-corrected block means to
absolute finite-aperture transmittance. Such a reference is not
reported for the 1-$\mu$m sample; its results are directional
ratios. In the analyzer experiment, two orthogonal output
projections are set at the extrema of transmitted brightness.
These projections do not select pure incident TE and TM states.
The experiment measures reproducible finite-aperture transmission
under fixed extended illumination, rather than complete
angle-resolved diffraction efficiencies or hemispherical
transmittance.

\section{Growth model and Maxwell calculation}

The ballistic-growth model uses collimated incident flux at angle
$\alpha$ from the substrate normal, with its azimuth perpendicular
to the grating lines, unity sticking, line-of-sight occlusion, and
cellular growth. At $\alpha=45^\circ$ the deposited thickness
after $K$ growth steps is $\tau\approx0.0081\,Kd$. Within this
model the deposition-axis channel remains open asymptotically,
with open fraction saturating at $0.28$, while the straight-through
channel closes for $\tau\gtrsim0.26\,d$. These are predicted
geometric properties, not cross-sectional measurements of the
optical samples. The EDS comparison in Fig.~1(d) of the main text
passes the growth profile through a minimal emission response
with generation depth $0.4\,\mu$m.

The two-dimensional FDTD solver uses a Yee grid with 128 cells
per period ($\geq34$ cells per vacuum wavelength at
$d/\lambda=3.6$), one period with periodic lateral boundaries,
and convolutional perfectly matched layers. The reference
geometry has line height $0.5\,d$, duty cycle $0.5$, and a finite
substrate slab of thickness $0.375\,d$. The dielectric has
$\varepsilon=n^2$, $n=1.55$, and silver is treated as a perfect
conductor in the reference calculation. Broadband pulses enter
at normal incidence from the surrounding vacuum on each side.
TE ($E\parallel$ lines) and TM ($H\parallel$ lines) are computed
separately. The reported $T^{F,B}$ is the net transmitted flux
summed over all propagating orders and normalized to a vacuum
reference. For each direction,
$T_\mathrm{unpol}=(T_\mathrm{TE}+T_\mathrm{TM})/2$, and the
contrast is $C_\mathrm{unpol}=T^B_\mathrm{unpol}/T^F_\mathrm{unpol}$.
The solver reproduces the analytic slab Fabry--P\'erot response
to $0.6\%$ and conserves energy to $|T+R-1|\leq0.9\%$ on the
reference grids. The geometric-regime checks are detailed below.

Spectra are compared through the Gaussian band average
\begin{equation}
\langle T\rangle(x_0)=
\frac{\int T(x)\,e^{-(x-x_0)^2/(2\sigma^2)}\,dx}
{\int e^{-(x-x_0)^2/(2\sigma^2)}\,dx},
\qquad x=d/\lambda.
\label{eq:band}
\end{equation}
The widths are $\sigma=0.25$ in the diffraction regime, matching
the scalar wave-packet spectral width, and $\sigma=1.2$ in the
geometric regime. This smoothing characterizes the computed
spectra; it is not a convolution with measured LED spectra,
detector responsivity, or the experimental angular acceptance.

The scalar wave-packet comparison in Fig.~3 of the main text is
the isotropic massless limit of the Dirac engine cited there,
with silver represented as an infinite-mass mirror. It provides
a conceptual comparison of bulk propagation in asymmetric
geometries. The scalar and Maxwell models have different boundary
conditions, so quantitative optical predictions rely on Maxwell
electromagnetics.

\section{Material, profile, and deposition-angle predictions}

The Drude-silver calculation uses auxiliary-current FDTD with
the Ag plasma frequency and damping; slab reflectance is checked
against the analytic Fresnel response. For the 1-$\mu$m model,
the visible-band contrast remains within $\sim10\%$ of the
perfect-conductor values (TM $1.2$--$1.6$ across the measured
channels). It predicts direction-dependent absorption,
$A_\mathrm{back}\approx0.08$--$0.10$ and
$A_\mathrm{front}\approx0.02$--$0.03$, consistent with favored
transmission grazing the metal walls. The unpolarized spectrum
has its largest contrast in the red. In the reported calculation,
the perfect-conductor description degrades toward the plasma
regime ($d/\lambda\gtrsim3$ for $d=1\,\mu$m). These are computed
material effects; an experimental transmission/reflection/
absorption balance is not available.

A shallow-profile variant fitted to the Ag~L$\alpha$ lateral
profile of Fig.~1(d) of the main text uses line height $0.19\,d$,
duty cycle $0.4$, and $\tau=0.06\,d$. It retains the
substrate-favored sign with unpolarized contrast $1.03$--$1.09$.
In the calculated dose series, contrast rises monotonically from
$\approx0.9$ at $\tau\approx0.07\,d$ to $1.3$--$2.0$ at
$\tau\approx0.23\,d$, crossing unity near
$\tau\approx0.1\,d$. At smaller thickness the modeled overhangs
are less developed and the structure is weakly front-favored;
the predicted sign inversion at fixed deposition angle has not
been established by the reduced-dose measurement alone.

Raw TM spectra contain sharp Rayleigh--Wood features at the
in-medium thresholds $d/\lambda=m/n$ identified in Fig.~3(a)
of the main text. The visible range of the 1-$\mu$m grating
samples this diffraction regime. A wavelength-resolved
measurement would test the predicted spectral structure beyond
the existing RGB measurements.

\begin{figure}[tbp]
\centering
\includegraphics[width=8cm]{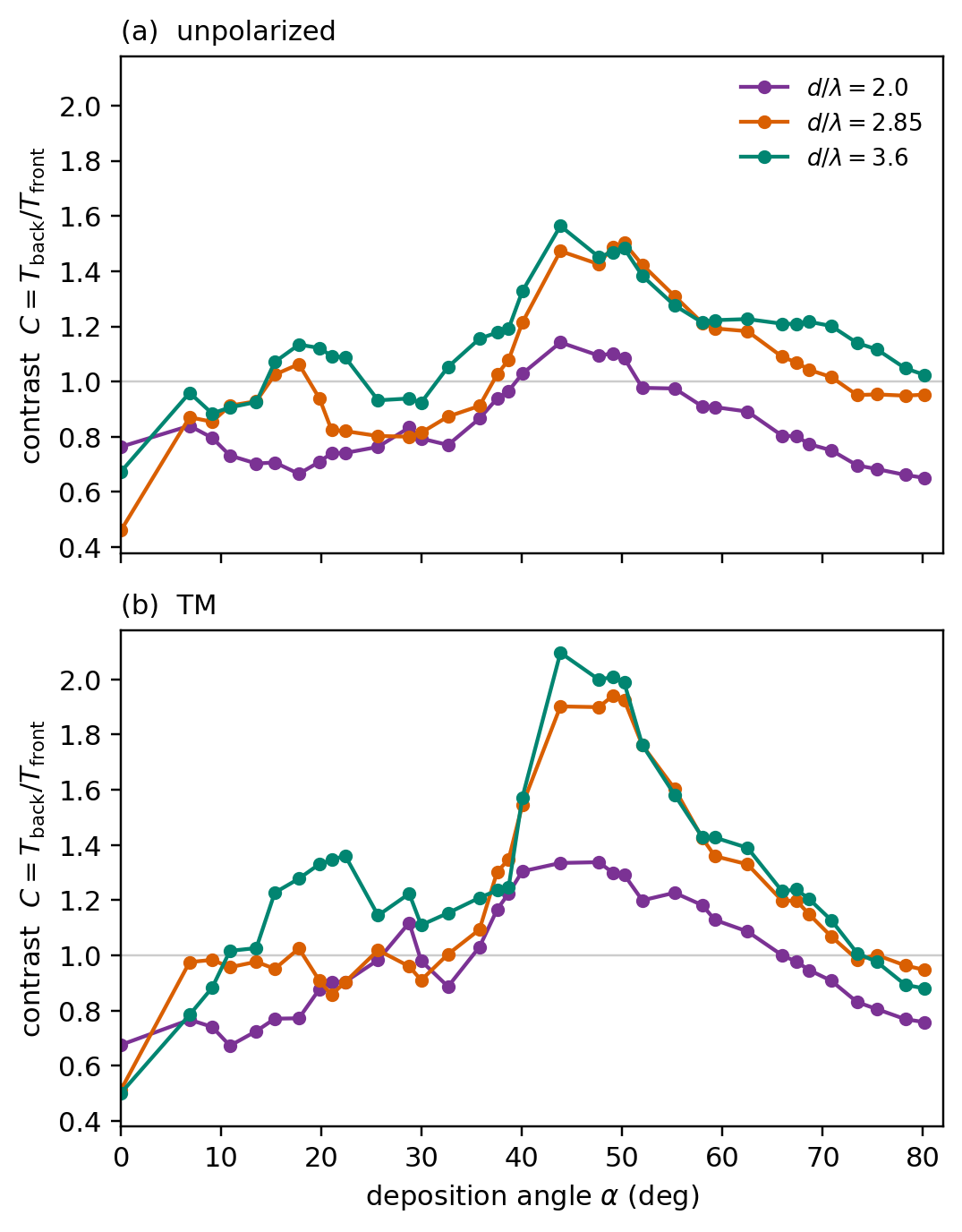}
\caption{\label{fig:angle}
\textbf{Predicted deposition-angle dependence (Maxwell, 35 angles).}
Band-averaged contrast versus deposition angle $\alpha$ for
(a)~unpolarized and (b)~TM illumination at three $d/\lambda$
(color), for ballistic-growth films at fixed source exposure
($K=20$; grazing angles intercept reduced projected flux and
receive less material). Circles mark the computed angles; lines
are guides to the eye.}
\end{figure}

The calculated angle scan of Fig.~\ref{fig:angle} covers 35 deposition
angles between $0^\circ$ and $80^\circ$ at fixed source exposure
($K=20$). Deposited material per substrate area decreases toward
grazing incidence with the projected flux. The calculated response
is nonmonotonic, with some wavelengths crossing $C=1$ more than
once: near-normal coating is front-favored ($C\approx0.7$ at
$d/\lambda=2$); low-angle substrate-favored lobes occur for
$d/\lambda=2.85$ and $3.6$; and the principal transition into the
broad high-angle substrate-favored region shifts to larger
$\alpha$ as $d/\lambda$ decreases. Contrast peaks near
$44^\circ$--$50^\circ$ and declines toward grazing incidence
with a wavelength-dependent sign. In a separate model comparison,
equal deposited volume at grazing angles seals the grooves,
producing opaque mirrors.

The near-normal prediction applies to continuous low-loss
coatings. The reported hard-wall and finite-skin-depth comparisons
retain resonant inter-layer transmission with $C=0.3$--$0.6$.
Absorption is expected to damp these resonant channels, requiring
a lossy-metal refinement of this low-angle comparison. Neither
the full deposition-angle dependence nor reversal of the
deposition azimuth has been measured in the present optical
experiment. The normal-incidence deposition control establishes
the absence of one-sided Ag redistribution; it is a morphology
control, not a measured optical angle scan.

\section{Numerical convergence of the geometric-regime grids}

\begin{figure}[htbp]
\includegraphics[width=0.62\textwidth]{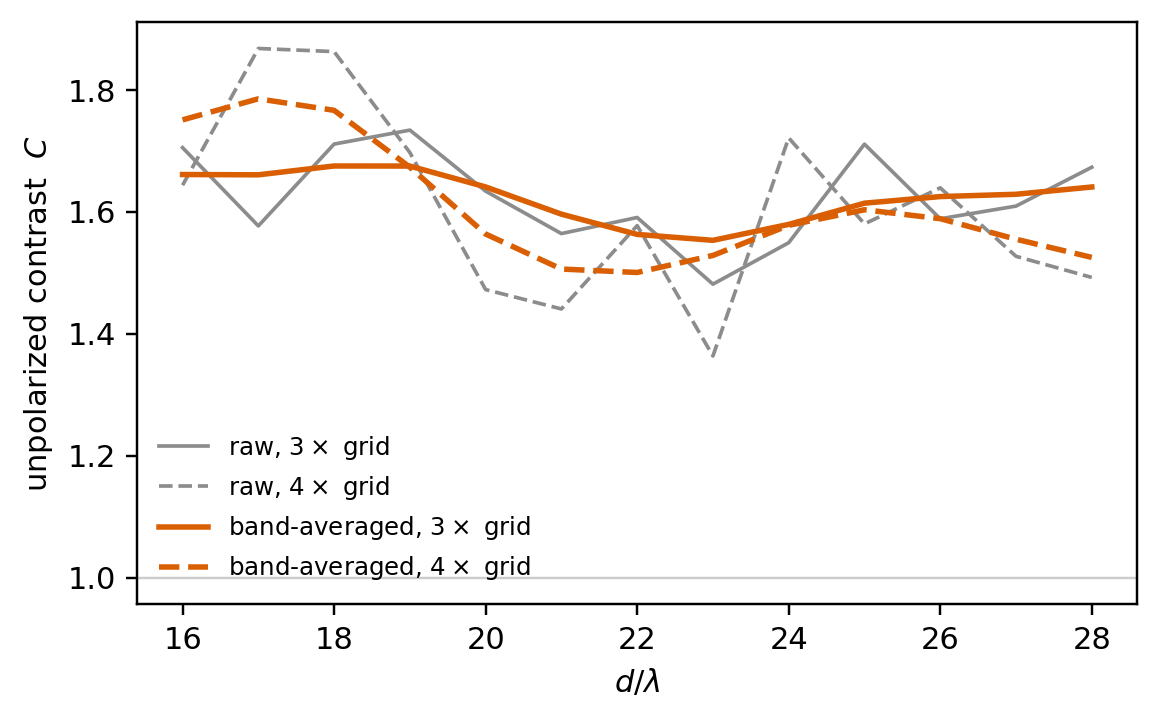}
\caption{\label{fig:conv}
\textbf{Grid convergence.} Unpolarized directional contrast on the
two finest grids at their common $d/\lambda$ points: raw grid
points (gray) and band-averaged curves ($\sigma_{d/\lambda} = 1.2$,
orange); $3\times$ grid solid, $4\times$ grid dashed.}
\end{figure}

The base finite-difference time-domain (FDTD) grid resolves one
grating period with $128$ cells.
The geometric-regime spectra of Fig.~3(b) of the main text are
computed on $3\times$ and $4\times$ upsampled versions of this
grid, which at $d/\lambda = 28$ resolve $13.7$ and $18.3$ cells
per vacuum wavelength. For the direct comparison the $3\times$
grid was evaluated at the same $d/\lambda = 16$--$28$ points as
the $4\times$ grid [Fig.~\ref{fig:conv}].

The raw spectra sample narrow Fabry--P\'erot-type resonances of
the metal--dielectric stack, and the resonance positions shift
slightly with grid resolution; point-by-point relative differences
between the two grids therefore reach $33\%$ in the individual
transmittances (mean $4$--$16\%$ over the overlap points) and
$16\%$ in the unpolarized contrast (mean $7\%$). The reported
quantities are the band-averaged curves ($\sigma_{d/\lambda} =
1.2$, bold in Fig.~3(b) of the main text); on these the two grids
agree to $7.6\%$ maximum and $3.8\%$ mean for the unpolarized
contrast ($10.8\%$/$5.7\%$ for the transverse-electric (TE)
polarization, $17.0\%$/$10.2\%$ for the transverse-magnetic (TM)
polarization), with
energy conserved to $|T + R - 1| \leq 1.8\%$ on both grids. The
band-averaged contrast values of the main text carry an
$\approx 8\%$ discretization uncertainty at the largest
$d/\lambda$.

\end{document}